\documentstyle[aps,prl,floats,graphicx,epsfig]{revtex}
\begin{document}
\draft
\twocolumn[\hsize\textwidth\columnwidth\hsize\csname @twocolumnfalse\endcsname

%
\title{3He and Universe parallelism.}
\author{G.E. Volovik  }

\address{  Low Temperature Laboratory, Helsinki
University of Technology, Box 2200, FIN-02015 HUT, Espoo, Finland\\
and\\ Landau Institute for Theoretical Physics, Moscow, Russia}

\date{\today} \maketitle
\

\begin{abstract}

We discuss topological properties of the ground state of spatially
homogeneous ensemble
of fermions. There are several classes of
topologically different fermionic vacua; in each case the momentum space
topology of the vacuum
determines the low-energy (infrared) properties of the fermionic energy
spectrum.  Among them
there is class of the gapless systems which is characterized by the
Fermi-hypersurface, which is
the topologically stable singularity. This class contains the conventional
Landau Fermi-liquid
and also the non-Landau Luttinger Fermi-liquid. Another important class of
gapless systems is
characterized by the topologically stable point nodes (Fermi points).
Superfluid
$^3$He-A and electroweak vacuum belong to this universality class. The
fermionic
quasiparticles (particles) in this class are chiral: close to the Fermi
points they are
left-handed or right-handed massless relativistic particles.  Since the
spectrum becomes
relativistic at low energy, the symmetry of the system is enhanced in the
low-energy edge. The
low-energy dynamics acquires local invariance, Lorentz invariance and
general covariance, which
become better and better when the energy decreases.   Interaction of the
fermions near the Fermi
point leads to collective bosonic modes, which look like effective gauge
and gravitational
fields. Since the vacuum of superfluid $^3$He-A and electroweak vacuum are
topologically
similar, we can use  $^3$He-A for  simulation of many phenomena in high
energy physics,
including axial anomaly. $^3$He-A textures induce a nontrivial effective
metrics of the space,
where the free quasiparticles move along geodesics. With $^3$He-A one can
simulate event
horizons, Hawking radiation, rotating vacuum, conical space, etc.
\end{abstract}
\
]

\newpage
\tableofcontents

\section{Introduction}

The physical vacuum is a complicated condensed matter
\cite{Hu,Wilczek,Jegerlehner,Jackiw}.
At the moment we know only the low-energy properties of this substance,
i.e. the properties at
energies much smaller than the Planck energy scale, $E\ll E_{\rm
P}=\sqrt{\hbar c^5/G}$, where
$G$ is Newton's constant. We know that at the low-energy edge our vacuum
has many different
symmetries: $U(1)$ and $SU(3)$ gauge symmetries, Lorentz invariance,
general coordinate
invariance, and dicrete CPT symmetry. With increasing energy more elements
of symmetry are
added: $SU(2)$ symmetry of weak interactions, probably the GUT symmetry and
even supersymmetry.
When the temperature dicreases such symmetries become spontaneously broken.
This is the
traditional point of view, which is supported by the observation of the
similar symmetry
breaking in different (many-body) condensed matter systems: superfluids,
superconductors, magnets,
liquid and ordinary crystals.

However there is another and actually opposite point of view: all symmetries
known in the Universe spontaneously (but without any phase transition)
appear at the low-energy
corner\cite{Chadha}. They become more and more pronounced the lower the
energy. The symmetries
disappear at higher energies when the Planck energy is approached. This
conjecture is also
supported by the condensed matter analogy.

Here we discuss the topological origin of such enhanced symmetry in the
infrared
limit. It is caused by the so-called topologically stable Fermi points in the
spectrum of the fermionic system. We show that if such points exist in quantum
condensed matter, then at low enough temperature this system exhibits fully or
partially the Lorentz invariance, general covariance, and gauge invariance.
Moreover
the collective modes which describe such a system at low $T$ all are
represented by
the chiral relativistic fermions, gauge fields, and gravity.  All are  the
low-energy
phenomena, which are absent at higher energies, where the condensed matter
has only a
very limited set of the global symmetries.

From the second point of view some directions in physics look artificial.
In particular,
since the gravity exists only as the infrared phenomenon one should not
quantize gravity:  Only
low-energy gravitons can be quantized \cite{Hu96}. The same concerns the
multidimensional string
theories,  which also give rise to gravitation in the infrared limit. The
Fermi point
mechanism does not require a high dimensionality for the space-time:  The
topologically stable
Fermi point is just a property of the conventional 3+1 dimensional space-time.

\section{Manifolds of zeroes.}

The Fermi point is a particular case of the topologically stable manifolds
of zeroes in the
fermionic spectrum. Let us start with the simplest such manifold -- the
Fermi surface.

\subsection{Fermi surface as topological object}

\begin{figure}[!!!t]
\begin{center}
\leavevmode
\epsfig{file=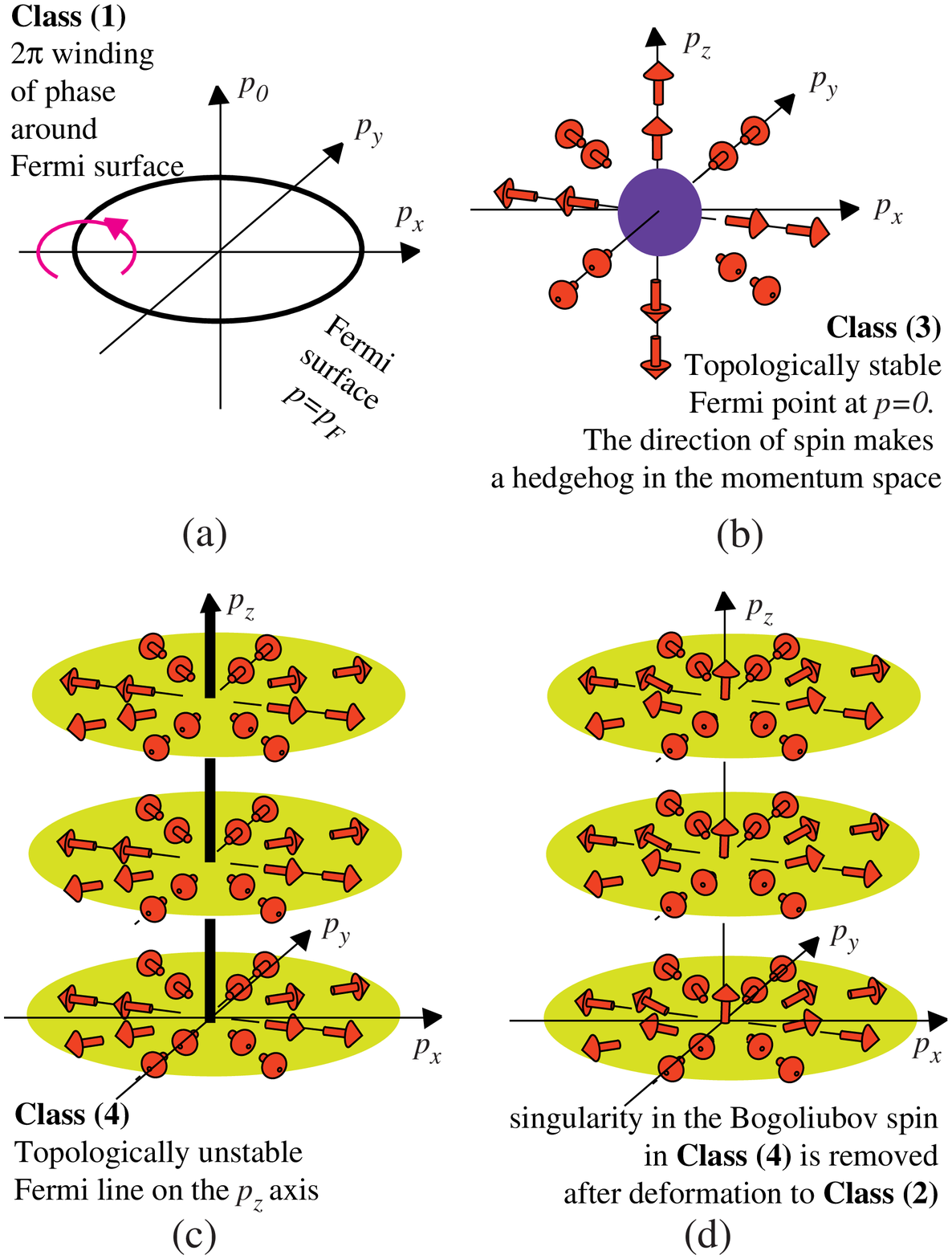,width=0.9\linewidth}
\caption[FermiSurfaceAsVortex]
    {(a) The Fermi surface  in the 2+1 momentum space. For simplicity the
$p_z$ coordinate is suppressed so that the Fermi surface is the line
$(p_0=0, p=p_F)$ in the 2+1   space $(p_0,p_x,p_y)$.  This line is a
singularity, which is similar to a vortex in a real 3D-space: The phase of
the propagator $G=(ip_0- (p_x^2 +p_y^2 -p_F^2)/2m)^{-1}$ changes by
$2\pi$ around the line in the momentum space in the same manner as the
phase of the order parameter changes by $2\pi $ around a vortex in the
real space. In general case the winding number $N_1$ -- the topological
invariant of the Fermi surface -- is given by  Eq.(\ref{InvariantForFS}).
(b) Fermi point at ${\bf p}=0$ in the 3D momentum space
$(p_x,p_y,p_z)$. At this point the particle energy $E=cp$ is zero.  A
right-handed particle is considered with its spin  parallel to the
momentum ${\bf p}$, {\it i.e.} ${\bf s}({\bf p})= (1/2){\bf p}/p$. The
spin makes a hedgehog in the momentum space, which is topologically
stable. The topological invariant of the Fermi point is $N_3=1$,
where $N_3$ is given by Eq.(\ref{TopInvariant}). (c) Fermi line --
topologically unstable manifold of zeroes -- is shown in the 3D momentum
space
$(p_x,p_y,p_z)$. The (Bogoliubov) spin (arrows) is confined into the
$(p_x,p_y)$
plane and has a singularity on the $p_z$ axis. (d) This singularity can be
removed bya continuous
transformation. The spin escapes into a third dimension ($p_z$) and becomes
well defined on the
$p_z$ axis. As a result, the quasiparticle spectrum becomes fully gapped
(the "relativistic"
fermion acquires the mass).}
\label{FermiSurfaceAsVortex}
\end{center}
\end{figure}

\subsubsection{Topological stability of Fermi surface}

The Fermi surface appears in the noninteracting Fermi gas, where the energy
spectrum of fermions
is
\begin{equation}
E(p)={p^2\over 2m}-\mu~,
\label{FermiGasEnergySpectrum}
\end{equation}
and  $\mu >0$ is the chemical potential.
The Fermi surface bounds the volume
in the momentum space where the energy is negative, $E(p)<0$, and where the
particle states
are all occupied at
$T=0$. In this isotropic model the Fermi surface is a sphere of radius
$p_F=\sqrt{2m\mu}$.

It is important that the Fermi surface survives even if
interactions between particles are introduced. Such stability of Fermi
surface comes from the
topological property of the Feynman quantum mechanical
propagator -- the one-particle Green's function
\begin{equation}
{\cal G} =(z-{\cal H})^{-1}~.
\label{Propagator1}
\end{equation}
Let us write the propagator for a given momentum ${\bf p}$ and for the
imaginary frequency,
$z=ip_0$. The imaginary frequency is introduced to avoid the conventional
singularity of the propagator "on the mass shell", i.e. at $z=E(p)$. For
noninteracting particles
the propagator has the form
\begin{equation}
G={1\over ip_0 -E(p)}~.
\label{Propagator2}
\end{equation}
Obviously there is still a singularity:  On
the 2D hypersurface
$(p_0=0, p=p_F)$ in the 4-dimensional space $(p_0, {\bf p})$ the propagator
is not well
defined. This singularity is stable, i.e. cannot be eliminated by small
perturbations. The
reason is that the phase $\Phi$ of the Green's funtion $G=|G|e^{i\Phi}$
changes by
$2\pi$ around the path embracing this 2D hypersurface in the 4D-space (see Fig.
\ref{FermiSurfaceAsVortex}). The phase winding number $N=1$ cannot change
continuously, that is
why it is robust towards any perturbation. Thus the singularity of the
Green's funtion on the
2D-surface in the momentum space is preserved, even when interactions
between particles are
introduced.

Exactly the same topological conservation of the winding
number leads to the stability of the quantized vortex in superfluids and
superconductors, the
only difference being that, in the case of vortices, the phase winding
occurs in the real space,
instead of the momentum space. The complex order parameter
$\Psi=|\Psi|e^{i\Phi}$ changes by
$2\pi N$ around the path embracing vortex line in 3D space or vortex sheet
in 3+1 space. The
connection between the topology in real space and the
topology in momentum space  is, in fact, even deeper (see {\it  e.g.}
Ref.\cite{VolovikMineev1982}). If the order parameter depends on
space-time, the
propagator in semiclassical aproximation depends both on 4-momentum and on
space-time
coordinates $G(p_0,{\bf p}, t, {\bf r})$. The topology in the 4+4
dimensional space describes:
(i) The momentum space topology in the homogeneous system; (ii) The defects
of the order
parameter in real space; and (iii) Topology of the energy spectrum within the
topological defects \cite{Grinevich1988}.
\begin{figure}[!!!t]
\begin{center}
\leavevmode
\epsfig{file=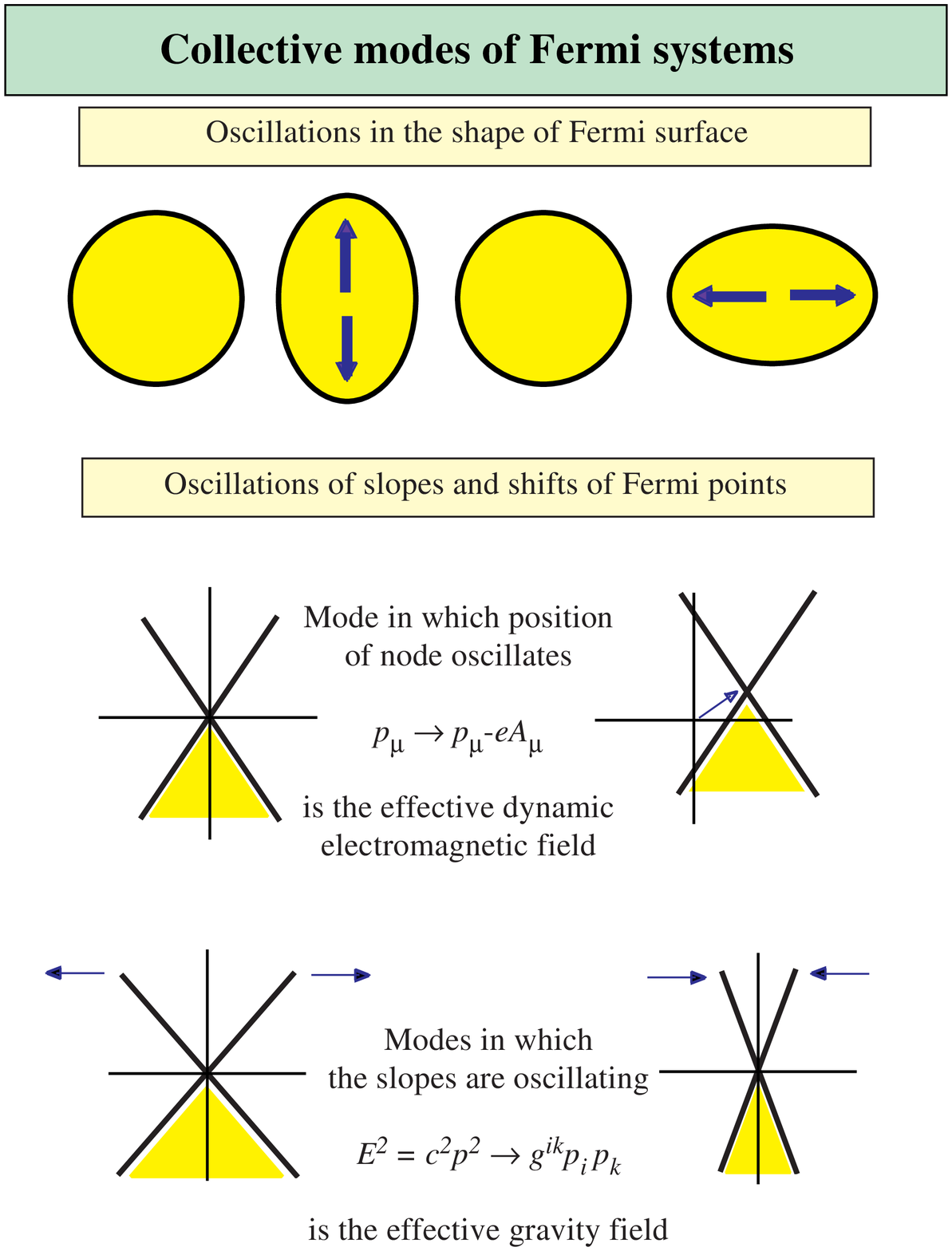,width=0.9\linewidth}
\caption[CollectiveModes]
    {Collective modes of fermionic systems with Fermi surface and Fermi point.}
\label{CollectiveModes}
\end{center}
\end{figure}

In the more complicated cases, when the Green's function is the matrix with
spin and band
indices,  the phase of the Green's function becomes meaningless. In this
case one should use  a
general analytic expression for the integer topological invariant which is
responsible for the
stability of the Fermi surface:
\begin{equation}
N_1={\bf Tr}~\oint_C {dl\over 2\pi i} {\cal G}(p_0,p)\partial_l {\cal
G}^{-1}(p_0,p)~.
\label{InvariantForFS}
\end{equation}
Here the integral is taken over an arbitrary contour $C$ in the momentum
space  $({\bf p},p_0)$,
which encloses the Fermi  hypersurface; and ${\bf Tr}$ is the trace over
the spin and band
indices.

\subsubsection{Landau Fermi liquid}

The topological class of systems with Fermi surface is rather broad. In
particular it contains
the conventional Landau Fermi-liquid, in which the propagator preserves the
pole. Close to the
pole the propagator is
\begin{equation}
G={Z\over ip_0 -v_F(p-p_F)}~.
\label{Propagator2}
\end{equation}
Evidently the residue $Z\neq 1$ does not change the  topological invariant
for the propagator,
Eq.(\ref{InvariantForFS}), wihch remains $N_1=1$. This is essential for the
Landau theory of an interacting Fermi liquid; it confirms the assumption
that in Fermi liquids
the spectrum of quasiparticles at low energy is similar to that of
particles in a Fermi gas. It
is also important for the consideration of the bosonic collective modes of
the Landau
Fermi-liquid. The interaction between the fermions cannot not change the
topology of the
fermionic spectrum,  but it produces the effective field acting on a given
particle by the other
moving particles. This effective field cannot destroy the  Fermi surface
owing to its topological
stability, but it can locally shift the  position of the  Fermi surface.
Therefore a collective
motion of the particles is seen by an individual quasiparticle as a
dynamical mode of the  Fermi
surface. These bosonic oscillative modes are known as the different
harmonics of the zero sound.
An example is shown in the upper part of Fig.\ref{CollectiveModes}.

Note that the Fermi hypersurface exists for any spatial dimension. In the
2+1 dimension the
Fermi hypersurface is a line -- the vortex line in the 2+1 momentum space
as is shown in Fig.
\ref{FermiSurfaceAsVortex}(a).

\subsubsection{Non-Landau Fermi liquids.}

In the
1+1 dimension the Green's function looses its pole, but nevertheless the
Fermi surface is still
there \cite{NewClass,Blagoev}. In 1+1 dimension the Landau Fermi liquid
transforms to
another states within the same topological class. Example is provided by the
Luttinger liquid. Close to the Fermi surface the Green's function for the
Luttinger
liquid can be approximated as (see
\cite{Wen,NewClass,LuttingerLiquidReview})
\begin{eqnarray}
\nonumber G(z,p)\sim \\
(ip_0 -v_1\tilde p)^{g-1\over 2}(ip_0+v_1\tilde p)^{g\over
2}(ip_0 -v_2\tilde p)^{g-1\over 2}(ip_0+v_2\tilde p)^{g\over 2}
\label{Propagator3}
\end{eqnarray}
where $v_1$ and $v_2$ correspond to Fermi velocities of spinons and holons and
$\tilde p=p-p_F$. The above equation is not exact but reproduces the
topology of
the Green's function in Luttinger Fermi liquid. If
$g\neq 0$ and
$v_1\neq v_2$, the singularity in the $(\tilde p, z=ip_0)$ momentum space
occurs
on the Fermi surface, i.e. at  $(p_0=0,
~\tilde p=0)$.  The  topological
invariant in Eq.(\ref{InvariantFor FS}) remains the same
$N_1=1$, as for the conventional Landau Fermi-liquid. The difference from
Landau Fermi liquid occurs only at real frequency $z$: The quasiparticle
pole is
absent and one has the branch cut singularities instead of the mass shell,
so that the
quasiparticles are not well defined. The population of the particles has no
jump
on the Fermi surface, but has a power-law singularity in the derivative
\cite{Blagoev}.

Another example of the non-Landau Fermi liquid is the  Fermi liquid with
exponential
behavior of the residue \cite{Yakovenko}. It also has the Fermi surface
with the same topological invariant, but the singularity at the Fermi
surface is
exponentially weak.

\subsubsection{Superconducting transition: from pole to zero}

In this section we do not consider the bifurcations which lead to the
overall reconstruction of
the particle spectrum, which occur, say, during the transition to
superconducting state. We assume
that the temperature, though is small compared to the "Planck" scale
$E_{\rm P}=\mu$, is still
above the superfluid transition: $T_c<T\ll E_{\rm P}$. But even in the
superconducting region
$T<T_c$ the winding number for the conventional Green's function is
preserved: the invariant in
Eq.(\ref{InvariantFor FS}) is again  $N_1=1$. Only instead of pole at
$p=p_F$  in the Green's function of the normal state, one has zero at
$p=p_F$ in the
superconducting state:
\begin{equation}
G=-{ip_0 +M(p)\over p_0^2+M^2(p) +\Delta^2}~, ~M({\bf p})={p^2\over
2m}-\mu\approx v_F(p-p_F)
\label{Propagator3}
\end{equation}
where $\Delta$ is the gap in the quasiparticle spectrum. This demonstrates
the stability of the
singularity in the Green's function, though the Fermi surface as the
manifold of zeroes in the
energy spectrum disappears in the superconductor.

\subsection{Fully gapped systems.}

Although the systems we have discussed contain fermionic and bosonic
quantum fields, this is
not the relativistic quantum field theory which we need for the simulation
of quantum vacuum:
There is no Lorentz invariance and the oscillations of the Fermi surface do
not resemble the
gauge field even remotely. The situation is somewhat better for superfluids
and superconductors
with fully gapped spectra; examples, which provide useful analogies with
Dirac fermions and spontaneously broken symmetry in quantum fields, are
conventional
superconductors
\cite{Nambu} and superfluid $^3$He-B \cite{ColorSuperfluidity}. In
$^3$He-B the Hamiltonian of
free Bogoliubov quasiparticles is the $4\times 4$ matrix:
\begin{eqnarray}
{\cal H}=\left( \matrix {M({\bf p}) & c{\bf
\vec\sigma}\cdot{\bf p}  \cr c{\bf \vec\sigma}\cdot{\bf p}  & -M({\bf p})\cr }
\right) =
{\bf \tau}_3 M({\bf p})+c{\bf \tau}_1 {\bf \vec\sigma}\cdot{\bf p}~,\\
 c={\Delta \over p_F}~~~,~~~{\cal H}^2=E^2=M^2({\bf p}) + \Delta^2~,~
\label{PropagatorFullyGapped}
\end{eqnarray}
where the Pauli $2\times 2$   matrices ${\bf \vec\sigma}$ describe  the
conventional spin of
fermions and $2\times 2$   matrices
${\bf\vec\tau}$ describe the Bogoliubov isospin in the particle-hole space.
The  Bogoliubov  Hamiltonian asymptotically approaches
the Dirac one in the low $p$ limit. This is however is not the low energy
limit, which for
the typical $^3$He-B parameters occurs for $p$ close to $p_F$. The low $p$
and low $E$
limits coincide at large negative $\mu$, which is not the case in $^3$He-B.
Nevertheless the $^3$He-B also serves as a
model system for simulations of many phenomena in particle physics and
cosmology including
experimental verification \cite{MiniBigBang} of the Kibble mechanism
describing formation of
cosmic strings in the early Universe \cite{Kibble}.

Note that even for the fully gapped systems there can exist the
momentum-space topological
invariants, which characterize the vacuum states. Typically this occurs in
2+1 systems,  e.g. in
the 2D electron system exhbiting quantum Hall effect \cite{Ishikawa}; in
thin film of $^3$He-A
(see \cite{VolovikYakovenko} and Sec.9 of Ref.\cite{exotic}); and in the
2D superconductors with broken time reversal symmetry
\cite{VolovikEdgeStates}. The quantum
(Lifshitz) transition between the states with different topological
invariants occurs through the
intermediate gapless regime
\cite{exotic}.

\subsection{Fermi point}

\subsubsection{Systems with Fermi points: $^3$He-A and Standard Model}

Now we proceed to the topological class which most fully exhibits the
fundamental
properties needed for a realization of the relativistic quantum fields,
analogous to those in
particle physics and gravity.

It is the class of systems, whose representatives are superfluid
$^3$He-A and the vacuum of relativistic left-handed and right-handed chiral
fermions (another
example of this class in condensed matter has been discussed for gapless
semiconductors
\cite{Abrikosov}). This class is
characterized by points in the momentum space where the (quasi)particle
energy is zero. In
particle physics the energy spectrum $E({\bf p})=cp$ is characteristic of
the massless
neutrino (or any other chiral lepton or quark in the Standard Model) with
$c$ being the speed of
light. The energy of a neutrino is zero at point ${\bf p}=0$ in the 3D
momentum space. The
Hamiltonian for the neutrino -- the massless spin-1/2 particle -- is a
$2\times 2$ matrix
\begin{equation}
 {\cal H}=\pm
c\vec \sigma\cdot {\bf p}
\label{Neutrino}
\end{equation}
 which is expressed in terms of the Pauli spin matrices $\vec
\sigma$. The sign $+$ is for a right-handed particle  and $-$  for a
left-handed one:
the spin of the particle is oriented along or opposite to its momentum,
respectively.

The Bogoliubov matrix for the  $^3$He-A fermions is
\begin{equation}
{\cal H}={\bf \tau}_3 M({\bf p})~+~c({\bf \vec\sigma}\cdot\hat d)({\bf \tau}_1
{\bf e}^{(1)}\cdot {\bf p} - {\bf \tau}_2   {\bf e}^{(2)}\cdot {\bf p})~,
\label{APhaseFermions1}
\end{equation}
where ${\bf e}^{(1)}$ and $  {\bf e}^{(2)}$ are real vectors, which in
equilibrium or
ground state are unit orthogonal vectors. The energy of the fermions
\begin{equation}
E^2_{{\bf p}}=M^2({\bf p}) +
c^2 ({\bf p}\times \hat {\bf l})^2 ~,  ~\hat {\bf l}\equiv {\bf e}^{(3)}={\bf
e}^{(1)}\times  {\bf e}^{(2)}~,
\label{APhaseFermions2}
\end{equation}
is zero at 2 points, at ${\bf p}=\pm p_F\hat{\bf l}$.

\subsubsection{Topological invariant for Fermi point}

Let us show that zeroes in the spectrum of the chiral fermions and in the
spectrum of  $^3$He-A
fermions are topologically stable and described by the same topological
invariant.
Let us again consider the propagator of the particle
${\cal G} =(ip_0-{\cal H})^{-1}$ on the imaginary frequency axis, $z=ip_0$.
One can see that this
propagator still has a singularity, which is now not on the surface but at
the points in  the 4D
momentum space:
$(p_0=0,{\bf p}=0)$ and $(p_0=0,\pm p_F\hat {\bf l})$ for chiral fermions
and  $^3$He-A
fermions respectively. An integer topological invariant which supports the
stability
of the point zeroes is expressed in terms of the propagator in the
following way
\cite{exotic}:
\begin{equation}
N_3 = {1\over{24\pi^2}}e_{\mu\nu\lambda\gamma}~
{\bf tr}\int_{\sigma}~  dS^{\gamma}
~ {\cal G}\partial_{p_\mu} {\cal G}^{-1}
{\cal G}\partial_{p_\nu} {\cal G}^{-1} {\cal G}\partial_{p_\lambda}  {\cal
G}^{-1}~,
\label{TopInvariant}
\end{equation}
where $\sigma$ is the 3-dimensional surface embracing the
point node in the 4-momentum space.

For the chiral fermions in Eq.(\ref{Neutrino}) this
invariant is $N_3=\pm 1$, where the sign is determined by the chirality of
the fermion.
The meaning of this topological invariant can be easily visualized. Let
us consider the behavior of the particle spin ${\bf s}({\bf p})$ as a
function of its
momentum ${\bf p}$ in the 3D-space   ${\bf p}=(p_x,p_y.p_z)$. For
right-handed particles
${\bf s}({\bf p})={\bf p}/2p$, while for left-handed ones ${\bf s}({\bf
p})=-  {\bf
p}/2p$. In both cases the spin distribution in the momentum space looks
like a hedgehog (see
Fig. \ref{FermiSurfaceAsVortex}b), whose spines are represented by spins:
spines point outward
for the right-handed particle and inward for the left-handed one. In the
3D-space the hedgehog
is topologically stable.

\subsubsection{Spin from isospin.}

 For the $^3$He-A fermions  in Eq.(\ref{APhaseFermions1}) $N_3=\mp 2$,
where the sign is
determined by the position of the node,
${\bf p}=\pm p_F\hat {\bf l}$. The topological invariant is twice larger
because of the double
degeneracy of the Fermi point over the conventional spin of the $^3$He
atom. For each projection
of spin one has $N_3=\mp 1$.
Note that the Bogoliubov spin $\vec\tau$ in $^3$He-A plays the
same role as the conventional spin $\vec\sigma$ of chiral fermions in
Eq.(\ref{Neutrino}).
On the other hand the conventional spin of the $^3$He atom is responsible
for the degeneracy,
but not for chirality, and thus plays the part of the isospin (see also
Sec.\ref{degeneracy}).
This means that the origin of the spin  responsible for the chirality of
the (quasi)particle  is
fully determined by the matrix structure of the Fermi point. In this sense
there is no principle
difference between  spin and isospin: changing the matrix structure one can
convert
isospin to spin, while the topological charge of the Fermi point remains
invariant.

\subsection{Fermi line}

\subsubsection{Superconductivity in cuprates.}

The high-temperature superconductors in cuprates most probably
contain zeroes in their quasiparticle energy spectrum. The ARPES experiments
\cite{Ding} show that these are four lines in the 3D momentum
space where the quasiparticle energy is zero  or, equivalently, four point
zeroes in
the 2D CuO$_2$ planes.  The high-T superconductors thus belong to class of
systems with Fermi lines: the dimension $D$ of the manifold of zeroes is
$1$, which is
intermediate between a Fermi surface with
$D=2$ and a Fermi point with $D=0$.

The energy spectrum of quasiparticles near each of the 4 gap nodes can be
written as
\begin{equation}
{\cal H}=\tau_1 c^x(p_x-p_{x}^{(0)}) + \tau_3 c^y(p_y-p_{y}^{(0)})
~.
\label{Lines}
\end{equation}
 The "speeds" of light $c^x$
and $c^y$ are the "fundamental" characteristics determined by the
microscopic physics of the
cuprates. This means that the system belongs to the same class as 2+1 QFT
with massless
relativistic fermions.

\subsubsection{Scaling law near zeroes.}

As in the other two classes of fermionic systems with the gapless
quasiparticles, all  low-energy
(infrared) properties of cuprate superconductors are determined by zeroes.
In particular,
the density of the fermionic states is determined by the
dimension of the zeroes:
\begin{equation}
N(E)=\sum_{\bf p}\delta (E-E({\bf p})) \sim E^{2-D}~.
\label{DOSGeneral}
\end{equation}
Many low-temperature properties of these superconductors are obtained from
a simple
scaling arguments. For example, an external magnetic field $B$ has
dimension of $E^2$
and thus of $T^2$. At finite $B$, the density of states is nonzero even at
$E=0$.
Substituting
$B\sim E^2$ to Eq.(\ref{DOSGeneral}) one obtains  $N(0,B)\sim B^{(2-D)/ 2}$
and the
following scaling law for the heat capacity:
\begin{equation}
C(T,B)=B^{(2-D)/ 2}T~f({B\over T^2})~,
\label{HeatCapacity}
\end{equation}
where $f$ is some function with the known asymptotes (see \cite{Scaling}). An
experimental indication of such scaling with
$D=1$ was reported for YBa$_2$Cu$_3$O$_7$ in Ref.~\cite{Revaz}.

\subsubsection{Topological instability of Fermi line.}

The lines of zeroes generally have no  stability: There is no corresponding
$N_2$
invariant, which can support the topological stability. The singular line
in the momentum space
from which the spines (now the vector
$\vec\tau$) point outward (see Fig. \ref{FermiSurfaceAsVortex}c) can be
elmininated by the escape
of the  $\vec\tau$-vector to a third dimension. This can be accomplished by
an operation similar
to the folding of an umbrella  (see Fig. \ref{FermiSurfaceAsVortex}d).

Existence of the nodal lines can be prescribed, however, by the symmetry of
the ground
state. There are many  nontrivial classes  of
superconductors, whose symmetry supports  the existence of nodal lines in
symmetric
positions in the momentum space \cite{VolovikGorkov}. The symmetry
violating  perturbations,
such as impurities, an external magnetic field, etc., destroy the lines of
zeroes
\cite{Grinevich1988}. One could expect different types of transformations
of these lines of
zeroes which depend on the perturbation. Impurities, for example, can : (i)
produce the gap in
the fermionic spectrum \cite{Pokrovsky1995} (see
Fig.\ref{FermiSurfaceAsVortex}d), which corresponds  to appearance of mass
for the  2+1
relativistic fermions; (ii) lead to the finite density of states
\cite{Sun1995}, thus
transforming the system to  Class (1); (iii) produce zeroes of fractional
dimension, which means
that  the exponent in the density of states $N(E)\propto E^{2-D}$ is
non-integral
\cite{Nersesyan1994} and thus corresponds to a fractional $D$ of the
manifold of zeroes; and (iv)
lead to localization \cite{Lee1993}. An open question is: Can the quantum
fluctuations do the
same, in particular, can they change the effective dimension of the zeroes?

\section{Properties of system with Fermi points.}

\subsection{Relativistic massless chiral fermions.}

Close to the Fermi point  $p_{\mu}^{(0)}$ in the 4D space one can expand
the propagator in terms
of the deviations from this Fermi point, $p_\mu -p_{\mu}^{(0)}$. If the
Fermi point is not
degenerate, the general form of the propagator is
\begin{equation}
{\cal G}^{-1}=\tau^a e^\mu_a (p_{\mu} - p_{\mu}^{(0)})
~.
\label{GeneralPropagator}
\end{equation}
Here we returned back from the imaginary frequency axis to the real energy,
so that $z=E=-p_0$
instead of  $z=ip_0$; and $\tau^a=(1,\vec\tau)$. The quasiparticle spectrum
$E({\bf p}$ is given
by the poles of the propagator:
\begin{equation}
g^{\mu\nu}(p_{\mu} - p_{\mu}^{(0)})(p_{\nu} -
p_{\nu}^{(0)})=0~,~g^{\mu\nu}=\eta^{ab}e^\mu_a
e^\nu_b ~.
\label{GeneralEnergy}
\end{equation}
Thus in the vicinity of the Fermi point the massless quasiparticles are
described by the
Lorentzian metric
$g^{\mu\nu}$. It is most important that this is the general form of the
energy spectrum in the vicinity of any Fermi point, even if the underlying
Fermi system is not
Lorentz invariant; superfluid $^3$He-A is an example. The fermionic
spectrum necessarily
becomes Lorentz invariant near the Fermi point, {\it i.e.}. If one
applies this reasoning to our quantum vacuum, one may conclude that
possibly the observed Lorentz invariance of the physical laws is not
a fundamental but a low-energy property of the vacuum which results from
the existence of the topologically stable  Fermi points.

\subsection{Collective modes -- electromagnetic and gravitational fields.}

Let us consider the collective modes in such a system. The effective fields
acting on a given
particle due to interactions with other moving particles cannot destroy the
Fermi point. That
is why, under the inhomogeneous perturbation of the fermionic vacuum the
general form of
Eqs.(\ref{GeneralPropagator}-\ref{GeneralEnergy}) is preserved. However the
perturbations lead to
a local shift in the position of the Fermi point
$p_{\mu}^{(0)}$ in momentum space and to a local change of the vierbein
$e^\mu_a $ (which in
particular includes slopes of the energy spectrum (see Fig.
\ref{CollectiveModes}).  This means that the low-frequency collective modes
in such Fermi
liquids are the propagating collective oscillations of the positions of the
Fermi point and of
the slopes at the Fermi point. The former is felt by the right- or the
left-handed
quasiparticles as the dynamical gauge (electromagnetic) field, because the
main effect of the
electromagnetic field
$A_\mu=(A_0,{\bf A})$ is just the dynamical change in the position of zero
in the energy
spectrum: in the simplest case
$(E-eA_0)^2=c^2 ({\bf p}-e{\bf A})^2$.

The collective modes related to a local change of the vierbein $e^\mu_a $
correspond to the
dynamical gravitational field. The quasiparticles feel the inverse tensor
$g_{\mu\nu}$ as the metric of the effective space in which they move along
the geodesic curves
\begin{equation}
ds^2=g_{\mu\nu}dx^\mu dx^\nu
\label{InverseMetric}
\end{equation}
Therefore, the collective modes related to the slopes
play the part of the gravity field (see Fig. \ref{CollectiveModes}).

Thus near the Fermi point the quasiparticle is the chiral massless fermion
moving in the
effective dynamical electromagnetic and gravitational fields.

\subsection{Gauge invariance, general covariance, conformal invariance.}

Another important property which results from the above equation is that
the fermionic
propagator in Eq.(\ref{GeneralPropagator}) is gauge invariant and even
obeys the general
covariance near the Fermi point.  For example, the local phase
transformation of
the wave function of the fermion, $\Psi \rightarrow \Psi e^{ie\alpha({\bf
r},t)}$
can be compensated by the shift of the "electromagnetic" field $A_\mu
\rightarrow
A_\mu + \partial_\mu \alpha$. These attributes of the electromagnetic ($A_\mu$)
and gravitational ($g^{\mu\nu}$) fields also arise spontaneously  as the
low-energy
phenomena.

Now let us discuss the dynamics of collective bosonic modes, $A_\mu$ and
$g^{\mu\nu}$. Since these are the effective fields their motion equations
do not
necessarily obey gauge inariance and general covariance. However, in some
special
cases such symmetries can arise in the low energy corner. What are the
conditions
for that?

The effective Lagrangian for the collective modes is obtained by
integrating over the
vacuum fluctuations of the fermionic field.  This principle was used by
Sakharov and
Zeldovich to obtain an effective gravity
\cite{Sakharov} and effective electrodynamics
\cite{Zeldovich}, both arising from fluctuations of the fermionic vacuum.
If the main
contribution to the effective action comes from the vacuum fermions whose
momenta
${\bf p}$ are concentrated near the Fermi point, {\it i.e.} where the fermionic
spectrum is linear and thus obeys the ``Lorentz invariance'', the result of the
integration is necessarily invariant under gauge transformation, $A_\mu
\rightarrow
A_\mu + \partial_\mu \alpha$, and has a covariant form. The
obtained effective Lagrangian then gives the Maxwell equations for
$A_\mu$ and the Einstein equations for $g_{\mu\nu}$, so that the
propagating bosonic
collective modes do represent the gauge bosons and  gravitons.

Thus two
requirements must be fulfilled -- (i) the fermionic system has a Fermi
point and (ii)
the main physics is concentrated near this Fermi point. In this case the system
acquires at low energy all the properties of the modern quantum field
theory: chiral
fermions, quantum gauge fields, and gravity. All these ingredients are
actually
low-energy (infra-red) phenomena.

There is another important symmetry obeyed by massless relativistic Weyl
fermions,
the conformal invariance -- the invariance under transformation
$g_{\mu\nu}\rightarrow a({\bf r},t)g_{\mu\nu}$.  In the extreme limit when the
vacuum fermions are dominatingly relativistic, the effective action for
gravity must
be conformly invariant. Such gravity, the so-called Weyl gravity, is
a viable rival to Einstein gravity  in modern cosmology
\cite{Mannheim,Edery}:  The Weyl gravity (i) can explain the galactic
rotation curves
without dark matter; (ii) it reproduces the Schwarzschild solution at small
distances;
(iii) it can solve the cosmological constant problem, since the
cosmological constant
is forbidden if the conformal invariance is strongly obeyed; etc. (see
\cite{Mannheim2}).

\subsection{$^3$He-A: Gauge invariance but no general covariance.}

Let us consider what happens in a practical realization of systems with
Fermi points in condensed
matter -- in $^3$He-A.
Close to the gap nodes, {\it i.e.} at energies $E\ll\Delta$, where $\Delta$
is the maximal
value of the gap in $^3$He-A which plays the part of
the Planck energy, the quasiparticles do obey the relativistic equation
\begin{equation}
g^{\mu\nu}(p_\mu - e A_\mu)(p_\nu - e A_\nu)=0
\label{Rel}
\end{equation}
Here $e=\pm$ is the "electric charge" and
simultaneously the chirality of the quasiparticles.
Let us consider the simplest situation, when the  $^3$He-A is in its vacuum
manifold, which is
characterized by two unit mutually orthogonal vectors $ {\bf e}^{(1)}$ and
${\bf e}^{(2)}$, and
neglect the spin degeneracy of the Fermi point. When we consider the
low-energy collective modes,
these vectors are slowly changing in space-time. In this situation the
effective metric and
effective electromagnetic field are given by:
\begin{eqnarray}
{\bf A}=p_F{\hat{\bf l}}~, ~A_0=p_F {\bf v}_s\cdot {\hat{\bf
l}}~,\label{EMFieldInAPhase}\\
\nonumber g^{ik}= v_F^2 (\delta^{ik} - \hat l^i \hat l^k) +  c^2 \hat l^i
\hat l^k -v_s^iv_s^k~,\\
 g^{00}=-1~,~g^{0i}=v_s^i~,
\label{GravFieldInAPhase}
\end{eqnarray}
where ${\bf v}_s$ is the superfluid velocity given by $v_{si} =(\hbar/2m) {\bf
e}^{(1)}\nabla_i{\bf e}^{(2)}$.

From above equations it follows that the fields, which act on the
"relativistic" quasiparticles
as electromagnetic and gravitational fields, have very strange behavior.
For example, the same
texture of the ${\hat{\bf l}}$-vector is felt by  quasiparticles as the
effective magnetic field
${\bf B}=p_F
\vec\nabla\times{\hat{\bf l}}$ according to Eq.(\ref{EMFieldInAPhase})  and
simultaneously it
enters the metric according to Eq.(\ref{GravFieldInAPhase}). Such field
certainly cannot
be described by the Maxwell and Einstein
equations together.
Actually the gravitational and electromagnetic variables coincide in
$^3$He-A only
when we consider the vacuum manifold: Outside of this manifold they split.
$^3$He-A, as any other fermionic system with Fermi point, has enough number
of collective modes to provide the analogs for the independent
gravitational and
electromagnetic fields. But some of these modes are massive in $^3$He-A. For
example the gravitational waves correspond to the modes, which are
different from
the oscillations of the  ${\hat{\bf l}}$-vector. As distinct from the photons
(orbital waves -- propagating oscillations of the ${\hat{\bf l}}$-vector)  the
gravitons are massive \cite{exotic}.

All these troubles occur because in $^3$He-A the main contribution to the
effective action for bosonic fields come from the vacuum fermions at the
"Planck"
energy scale,   $E \sim
\Delta$. These fermions are far from the Fermi points and their spectrum is
nonlinear. That is why in general the effective action  is not symmetric.

\subsection{Zero charge effect and Maxwell equations.}

There are, however, exclusions, for example, the action
for the  ${\hat{\bf l}}$-field contains the term with the logarithmically
divergent factor $\ln
(\Delta/\omega)$ \cite{exotic}. It comes from the zero charge effect,
experienced by the vacuum of
the massless fermions for whom the ${\hat{\bf l}}$-field acts as
electromagnetic field. Due to its
logarithmic divergence this term is dominanting at low frequency $\omega$:
the lower the
frequency the larger is the contribution of the vacuum fermions from the
vicinity of the Fermi point and thus the more symmetric is the Lagrangian
for the ${\hat{\bf
l}}$-field. As a result, in the very low-energy limit, when the
non-logarithmic contributions can
be neglected, the effective Lagrangian for the
$A_\mu$ becomes gauge invariant and even obeys the general covariance:
\begin{equation}
L={{\sqrt{-g}}\over {24\pi^2} }~{\rm ln} \left
({\Delta^2\over \omega^2
}\right)~g^{\mu\nu}g^{\alpha\beta}F_{\mu\alpha}F_{\nu\beta} ~~,
\label{EMLagrang}
\end{equation}
where  $F_{\mu\nu}$ is the strength of the effective electromagnetic field
$A_\mu$ from
Eq.(\ref{EMFieldInAPhase}) and $g^{\mu\nu}$ is the effective gravitational
field from
Eq.(\ref{GravFieldInAPhase}). In this regime the $A_\mu$ field does obey
the Maxwell equations in
a curved space.

\subsection{Why $^3$He-A is not perfect.}

On the other hand the "Einstein" action for $g_{\mu\nu}$ is highly
contaminated by many
noncovariant terms, which come from the integration over the
"nonrelativistic" high energy
degrees of freedom at "Planck" scale. In this sense the $^3$He-A, with its
given physical
parameters,  is not a perfect model for quantum vacuum. To remove the
polluting noncovariant
terms, the integration must be spontaneously cut-off at energies much below
the "Planck" scale,
$E\ll \Delta$, for example, due to strong quasiparticle relaxation.

The main reason why $^3$He-A
is not a good substance, is that the Fermi points of the left particles,
i.e. at ${\bf
p}=+p_F{\hat{\bf l}}$, and the Fermi points of the right particles, i.e. at
${\bf
p}=-p_F{\hat{\bf l}}$, are far from each orher in equilibrium. The
"perfect" condensed matter
would be such where all the Fermi points are at the origin, at ${\bf
p}=0$, as it happens in the standard model. However, inspite of the absence
of general covariance
even in the low-energy corner, many different properties of the physical
vacuum with a Fermi
point, whose direct observation are still far from realization, can be
simulated in $^3$He-A. One
of them is the chiral anomaly.

\subsection{Chiral anomaly.}

The chiral anomaly is the phenomenon which allows the nucleation of the
fermionic charge from the
vacuum \cite{Adler1969,BellJackiw1969}. Such nucleation results from the
spectral flow of the
fermionic charge through the Fermi point to high energy. Since the flux in
the momentum space is
conserved, it can be equally calculated in the infrared or in the
ultraviolet limits. In $^3$He-A it is much easier to use the infrared
regime, where the fermions
obey all the "relativistic" symmetries. As a result one obtains the same
anomaly equation, which
has been derived by  Adler and by Bell and Jackiw for the relativistic
systems. The rate of
production of quasiparticle number
$n=n_R+n_L$ from the vacuum in applied electric and magnetic fields is
\begin{equation}
\partial_\mu J^\mu ={1\over {8\pi^2}} (e_R^2-e_L^2)F^{\mu\nu}F^{*}_{\mu\nu}~,
\label{ChargeParticlProduction}
\end{equation}
Here $n_R$ and $n_L$ is the density of the right and left quasiparticles;
$e_R$ and $e_L$ are
their charges; and $F^{*}_{\mu\nu}$ is the dual field strength. Note
that the above equation is fully "relativistic". This equation has been
verified in  $^3$He-A
experiments \cite{BevanNature,AxialAnomaly}, where the "magnetic" ${\bf B}=p_F
\vec\nabla\times{\hat{\bf l}}$  and "electric" ${\bf E}=p_F
\partial_t{\hat{\bf l}}$ fields have been simulated by the space and time
dependent ${\hat{\bf
l}}$-texture. In particle physics the only evidence of axial anomaly is
related to the decay of
the neutral pion
$\pi^0\rightarrow 2\gamma$, although the anomaly is much used in different
cosmological scenaria
explaining an excess of matter over antimatter in the Universe (see review
\cite{Trodden}).

\subsection{Degeneracy of Fermi point as the origin of the non-Abelian
gauge field.}
\label{degeneracy}

In $^3$He-A the Fermi point (say, at the north pole)
is  doubly degenerate owing to the ordinary spin $\vec\sigma$ of the $^3$He
atom. This means
that in equilibtium the two zeroes, each with the topological invariant
$N_3=-1$, are at the same
point in momentum space. Let us find out what can be the consequences of
the Fermi point
degeneracy. It is clear that the collective motion can split the Fermi
points: positions of the
two  points can oscillate separately. Moreover, since the propagator is now
the $4\times 4$
matrix there can be the cross terms. If we neglect the degrees of freedom
related to the
vierbein then the collective variables of the system with the doubly
degenerate Fermi point
enter the fermionic propagator as
\begin{equation}
{\cal G}^{-1}=\tau^a e^\mu_a (p_{\mu} - eA_{\mu} -
e\sigma_{\alpha}W^{\alpha}_\mu)
~.
\label{Nonabelian FieldPropagator}
\end{equation}
The new effective field $W^{\alpha}_\mu$ acts on the chiral quasiparticles
as  $SU(2)$
gauge field. Thus in this effective field theory the ordinary spin of the
$^3$He atoms
plays the part of the weak isospin \cite{VolovikVachaspati,exotic}. The
"weak" field
$W^{\alpha}_\mu$ is also dynamical and in the leading logarithmic order
obeys the Maxwell
(actually Yang-Mills) equations.

This implies that the higher
symmetry groups of our vacuum can in principle arise as a consequence of
the Fermi point
degeneracy. For example, the 4-fold degeneracy of the Fermi point (which
implies that
there are, say, 4 left-handed fermionic species,  or 2 left-handed + 2
right-handed) can produce
the $SU(4)$ gauge group. In particle
physics the collective modes related to the shift of the 4-momentum are
discussed in terms of the
"generalized covariant derivative"
\cite{Martin,Sogami}. In this theory the gauge fields, the Higgs fields,
and Yukawa interactions,
all are realized as shifts of positions of the degenerate Fermi point, with
degeneracy
corresponding to different quarks and leptons.

In the Eq.(\ref{Nonabelian FieldPropagator}) we did not take into account
that dynamically the
vierbein can also oscillate differently for each of the two elementary
Fermi points. As a
result the number of the collective modes could increase even more. This is
an interesting
problem which must be investigated in detail. If the degenerate Fermi point
mechanism has
really some connection to the dynamical origin of the non-Abelian gauge
fields, we must connect
the degeneracy of the Fermi point (number of the fermionic species) with
the symmetry group of the
gauge fields. Naive approach leads to extremely high symmetry group. That
is why there should be
some factors which can restrict the number of the gauge and other bosons.
For example there can be
some special discrete symmetry between the fermions of the degenerate
point, which restricts the
number of massless bosonic collective modes.  Another source of the
reduction of the number of the
effective field has been found by Chadha and Nielsen
\cite{Chadha}. They considered the massless electrodynamics with different
metric (vierbein) for
the left-handed and right-handed fermions. In this  model the Lorentz
invariance is violated.
They found that the two metrics converge to a single one as the energy is
lowered. Thus in the
low-energy corner the Lorentz invariance becomes better and better, and at
the same time the
number of independent bosonic modes decreases.

Since the connections between QFT in the standard model and in $^3$He-A has
been discussed
in the earlier publications \cite{exotic,VolovikVachaspati,AxialAnomaly},
we concentrate here on
some problems related to gravitational analogy.

\section{Black hole in  $^3$He-A film.}

\subsection{Gravity by motion of superfluids. Sonic black hole.}

As we have seen from Eq.(\ref{GravFieldInAPhase}) the gravitational field
can be simulated in
$^3$He-A by the motion of the liquid with the superfluid velocity ${\bf
v}_s$ and by the
${\hat{\bf l}}$-texture \cite{JacobsonVolovik}. In this Section we
consider the situation when the  ${\hat{\bf l}}$-vector is uniform and
thus does not produce the effective gravitational field, so that the
gravitational effects come only from the motion of the superfluid vacuu.
The propagation of fermions in the moving liquid obeys the same equation
as propagation of relativistic particles in the gravitational field. The
same happens for propagating sound waves in normal fluids
\cite{UnruhSonic,Visser1999} and  phonons in
superfluid
$^4$He. In the simplest case of the radial motion of superfluid
$^4$He, the effective metric is expressed in terms of the radial superfluid
velocity $v_s(r)$
as
\begin{equation}
 ds^2=-\left(c^2- v_s^2(r)\right)dt^2+ 2v_s(r)drdt+dr^2+r^2d\Omega^2~,
\label{SphericallySymmetricField}
\end{equation}
 where $c$ now is a speed of sound in $^4$He (phonon velocity).
For $^3$He-A fermions the spherically symmetric metric occurs if
${\hat{\bf l}}$-field is radial. Then it follows from
Eq.(\ref{GravFieldInAPhase})
\begin{equation}
 ds^2=-\left(c^2- v_s^2(r)\right)dt^2+ 2v_s(r)drdt+{c^2\over v_F^2}dr^2
+r^2d\Omega^2~,
\label{SphericallySymmetricField}
\end{equation}

Kinetic energy of superflow plays the part of the
gravitational potential: $ \Phi=- v_s^2(r)/2$. If one chooses the velocity
field corresponding to the potential of the point body of mass $M$
\begin{equation}
v_s^2(r) = -2\Phi={2GM \over r}\equiv c^2{r_h\over r}~,
\label{Schwarzschild}
\end{equation}
one obtains the Panlev\'e-Gullstrand form of
Schwarzschild geometry (see e.g.
ref.\cite{Visser1999}).
Here $r_h$ denotes the position of the event horizon, where the velocity
reaches the "speed of light"  $c$. If the fluid moves
towards the origin, the low-energy quasipartilcles are trapped behind
the horizon, since their speed $c$ with respect to the fluid is less than
the velocity of the fluid.

Such sonic black hole was first suggested by Unruh for ordinary liquid
\cite{UnruhSonic}.
However since all the known  normal liquids are classical, the most
interesting quantum effects
related to the horizon cannot be simulated in such flow. Also the geometry
is such that it
cannot be realized:  in such radial flow inward the liquid is accumulated
at the
origin, so that this sonic black hole cannot be stationary.  In the other
scenario a horizon
appears in moving solitons, if the velocity of the soliton exceeds the
local "speed of light"
\cite{JacobsonVolovik}. This scenario has the same drawback: in finite
system the motion of the
soliton cannot be supported for a long time.  In a draining bathtub
geometry suggested in
Ref.\cite{Visser1997} the fluid motion can be made constant in time. However
the friction of the liquid, which moves through the drain, is the main
source of
dissipation.  The
superfluidity of the liquid does not help much in this situation. Horizon does
not appear since the "superluminal" motion with respect to the boundaries of
the drain tube is not allowed: The superflow becomes unstable
and superfluidity collapses (see
\cite{KopninVolovik}). Let us suggest a scenario, in which this collapse is
avoided. The superfluid motion becomes quasi-stationary and exhibits the event
horizon; the life time of the flow state is determined  by Hawking radiation.

\subsection{Simulation of 2D black hole}

\begin{figure}[!!!t]
\begin{center}
\leavevmode
\epsfig{file=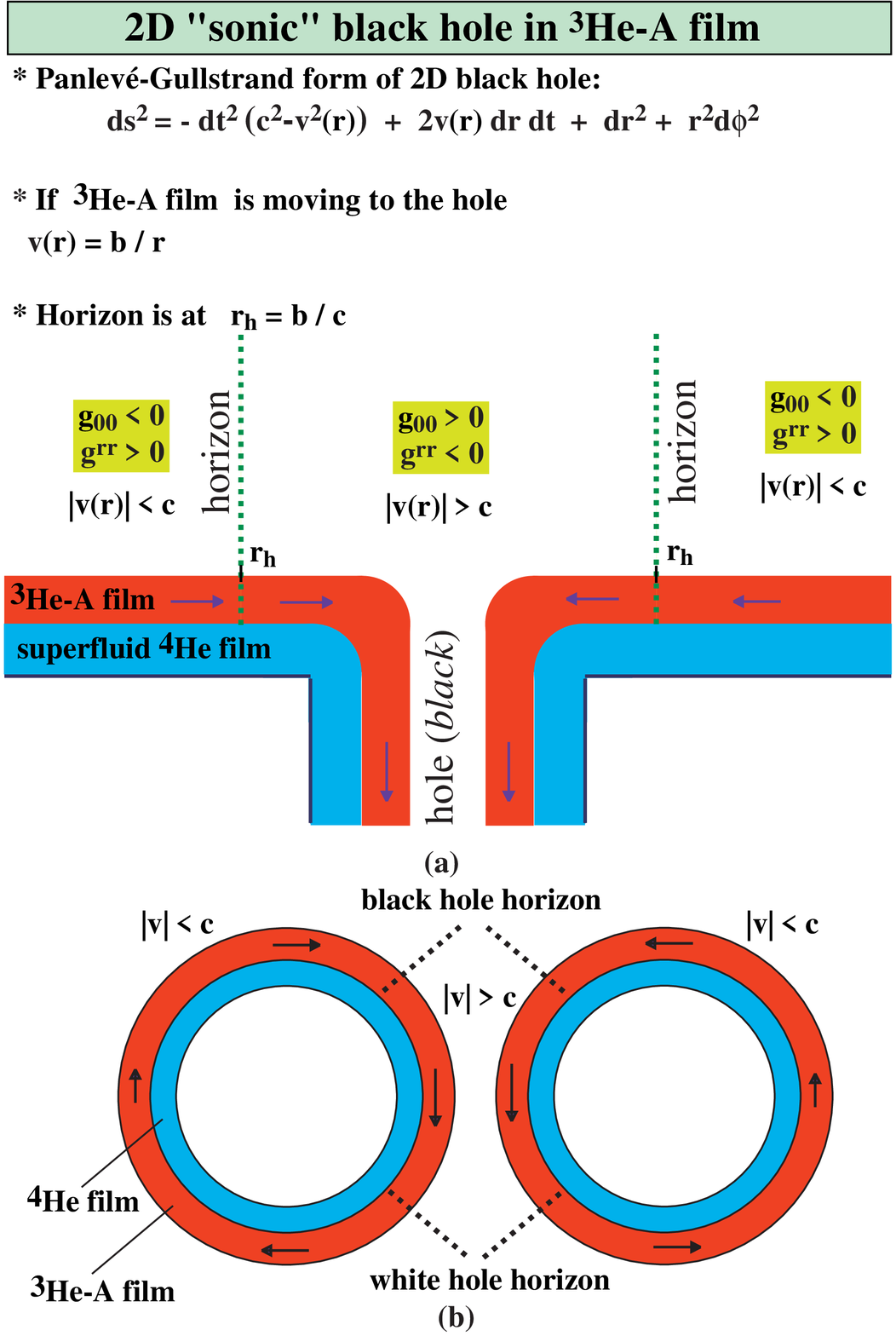,width=0.9\linewidth}
\caption[2DBH]
    {Similation of 2D black hole in thin $^3$He-A film. (a) Draining bathtub
geometry. (b) $^3$He-A film circulating on the top of the $^4$He film on
torus.}
\label{2DBH}
\end{center}
\end{figure}

 The stationary black hole can be realized
in the following geometry, which is the development of the  bathtub geometry of
Ref.\cite{Visser1997} (see Fig.
\ref{2DBH}(a)).  The superfluid
$^3$He-A film is moving towards the center of the disk, where it
escapes to the third dimension due to the orifice (hole). If the thickness
of the film is constant, the flow velocity increases towards the center as
$v_s(r)=b/r$ and at $r=r_h=b/c$ reaches the speed of light (now $r$
denotes the radial coordinate in the cylindrical system). Outside the
orifice the motion of the liquid is two dimensional and the effective
metric for the low-energy Bogoliubov quasiparticles is
\begin{equation}
 ds^2=-\left(c^2- v_s^2\right)dt^2+ 2v_sdrdt+
dr^2+r^2d\phi^2 +{c^2\over v_F^2}dz^2~.
\label{CylindricallySymmetricField}
\end{equation}
Here we took into account that the $\hat l$ vector in the film is fixed
along the normal to the film, so that the "speed of light" for
quasiparticles propagating along the film
$c \sim 3~$cm/sec is much smaller than the Fermi velocity $v_F$ which
corresponds to the "speed of light" for
quasiparticles propagating along the normal to the film. Note that $c$ is
much smaller than the speed of sound in $^3$He-A, that is why the motion of
fluid has no effect on the density of the liquid.

The important element of the construction in Fig.
\ref{2DBH} is that the moving superfluid  $^3$He-A film
is placed on the top of the superfluid $^4$He film. This is made to avoid the
interaction of the $^3$He-A film with the solid substrate.  The superfluid
$^4$He
film effectively screens the interaction and thus prevents the collapse of the
superfluid flow of $^3$He-A with "superluminal" velocity: Since the
interactions
with walls is removed the local observer moving with the superfluid
velocity cannot
detect the relative motion with respect to the wall.

The motion of the superfluid $^3$He-A with respect to superfluid $^4$He film is
not dangerous: The superfluid $^4$He is not excited even if $^3$He-A moves with
its superluminal velocity: $c$ for $^3$He-A is much smaller than the Landau
velocity for radiation of quasiparticles in superfluid $^4$He, which is about
$50$ m/sec. In this consideration we neglected the radiation of surface waves,
assuming that the thickness of $^4$He film is small enough.

Finally one can close the superflow by introducing the toroidal geometry in
Fig.
\ref{2DBH}(b), so that the superfluid condensate can circulate. In this case in
addition to the black hole horizon the white hole horizon appears on the path
where the superfluid $^3$He-A flows out from the orifice.

Since the  extrinsic
mechanism of the friction of $^3$He-A film -- the  scattering of
quasiparticles on
the roughness of substrate -- is abandoned, we can consider now intrinsic
mechanisms of dissipation. The most interesting one is the Hawking radiation
related to existence of a horizon.

\subsection{Vacuum in comoving and rest frames.}

Let us consider the simplest case of the 2D motion along the film in the
bathtub
geometry of Fig. \ref{2DBH}(a). This can be easily generalized to the motion in
the torus geometry.

\begin{figure}[!!!t]
\begin{center}
\leavevmode
\epsfig{file=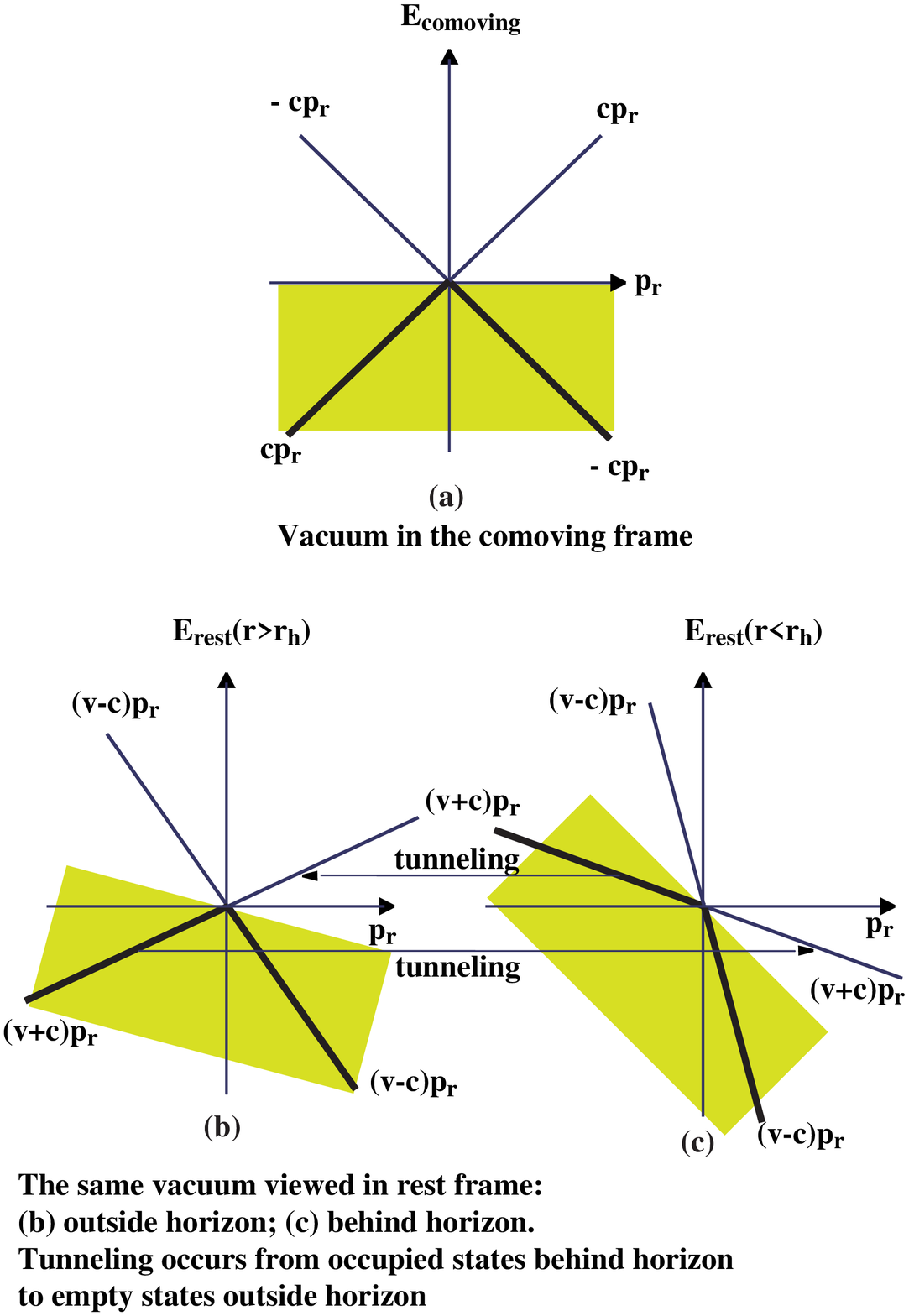,width=0.9\linewidth}
\caption[Vacua]
    {(a) Fermionic vacuum in the comoving frame. The states with $E_{\rm
com}<0$
are occupied (thick lines). The same vacuum viewed in the rest frame (b)
outside
horizon and (c) inside horizon. Behind the horizon the branch
$E_{\rm rest}=(v+c)p_r$ (for $p_\perp =0$) has inverse population as seen
in the
rest frame: the states with positive
enegry
$E_{\rm rest}>0$ are filled, while the states with
$E_{\rm rest}<0$ are empty. The tunneling across horizon from the occupied
states
to the empty states with the same energy gives rise to the Hawking
radiation from the horizon.}
\label{Vacua}
\end{center}
\end{figure}

There are two important reference frames:  (i) The
frame of the  observer, who is locally comoving with the superfluid vacuum.
In this frame the local superfluid velocity is zero,  ${\bf v}_s=0$, so
that the energy spectrum of the  Bogoliubov fermions in the place of the
observer is (here we assume a pure 2D motion along the film)
\begin{equation}
E_{\rm com}=\pm cp~.
\label{EnergyComovingFrame}
\end{equation}
In this geometry, in which the superflow velocity is confined in the plane
of the
film, the speed $c$ coincides with the Landau critical velocity of the
superfluid vacuum, $v_{\rm Landau}={\rm min} (|E_{\rm com}(p)|/p)$. The
vacuum as
determined by the comoving observer is shown on Fig.~\ref{Vacua}(a): fermions
occupy the negative energy levels in the Dirac sea (the states with the minus
sign in Eq.(\ref{EnergyComovingFrame}). It is the counterpart of the Minkowski
vacuum, which is however determined only locally: The comoving frame cannot be
determined globally. Moreover for the comowing observer, whose velocity changes
with time, the whole velocity field
${\bf v}_s({\bf r},t)$ of the superflow is time dependent. This does not
allow to determine the energy correctly.

(ii) The energy can be well defined in the laboratory frame (the
rest frame). In this frame the system is stationary, though is not static: The
effective metric does not depend on time,  so that the energy is conserved,
but this metric contains the mixed component
$g_{0i}=v_{si}$. The energy in the rest frame is
obtained from the local energy in the comoving frame by the Doppler
shift:
\begin{equation}
E_{\rm rest}=\pm cp + {\bf p}\cdot{\bf v}_s({\bf r})~.
\label{DopplerShift}
\end{equation}
 In case of the radial superflow ${\bf v}_s({\bf r})=\hat{\bf r} v_s(r)$
one has
\begin{equation}
E_{\rm rest}=\pm cp + p_rv_s(r)~.
\label{Energy}
\end{equation}

Figs.~\ref{Vacua}(b-c) show how the "Minkowski" vacuum of the comoving frame is
seen by the rest observer (note that the velocity is negative,
$v_s(r)<0$).   In the absense of horizon, or outside the horizon the local
vacuum does not change: the states which are occupied (empty) in the Minkowski
vacuum remain occupied (empty) in the rest frame vacuum (see
Fig.~\ref{Vacua}(b)).  In the presense of horizon behind which the velocity of
superflow exceeds the Landau critical velocity the situation changes:
Behind the
horizon the vacuum in the rest frame differs from that in the comoving
frame. Let
us for simplicity consider the states with zero transverse momentum
$p_\phi=0$ on the branch $E_{\rm rest}=(v_s(r)+c)p_r$ in the rest frame. If
the system is in  the Minkowski vacuum state (i.e. in the ground state as
viewed by comoving observer),  quasiparticles on this branch has reversed
distribution in the rest frame: the negative energy states are empty,
while the  positive energy states are occupied  (see
Fig.~\ref{Vacua}(c)). For this branch the particle distribution
corresponds to the negative temperature
$T=-0$ behind horizon.

Since the energy in the rest frame is a good quantum number, the fermions
can tunnel across the horizon from the occupied levels to the empty ones
with the
same energy. Thus if the system is initially in the Minkowski vacuum in the
comoving frame, the tunneling disturbs this vacuum state: Pairs of excitations
are created: the  quasiparticle, say, is created outside the horizon while its
partner -- the quasihole -- is created inside the horizon. This simulates the
Hawking radiation from the black hole.

\subsection{Hawking radiation}

\begin{figure}[!!!t]
\begin{center}
\leavevmode
\epsfig{file=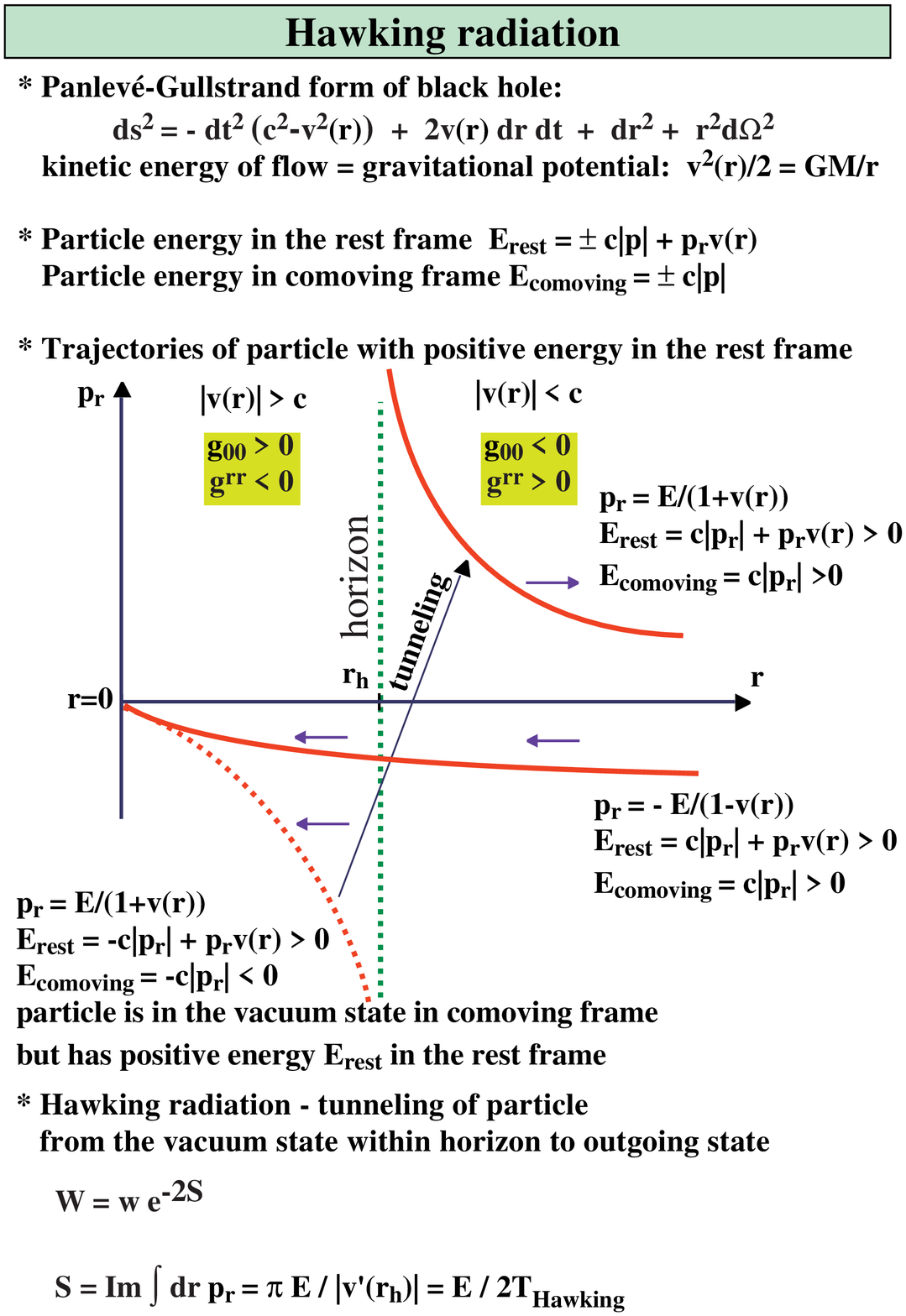,width=0.9\linewidth}
\caption[TunnelingTraj]
    {Tunneling from Minkowski vacuum within the horizon to the outgoing
mode. }
\label{TunnelingTraj}
\end{center}
\end{figure}

To estimate the tunneling rate in the semiclassical approximation, let us
consider
the classical trajectories $p_r(r)$ of particles, say, with positive energy,
$E_{\rm rest}>0$, for the simplest case when the transverse momentum
$p_\phi$ is
zero Fig.~\ref{TunnelingTraj}. The branch $E_{\rm rest}=(v_s(r)-c)p_r$
describes the  incoming particles with
$p_r<0$ which propagate through the horizon to the orifice (or to the
singularity at $r=0$, if the orifice is infinitely small) without any
singularity
at the horizon. The classical trajectories of these particles are
\begin{equation}
p_r(r)=-{E_{\rm rest}\over c-v_s(r)} <0 ~.
\label{IncomingTrajectory}
\end{equation}
The energies of these particles viewed by the comoving observer are also
positive:
$E_{\rm com}(r)= -cp_r(r) = E_{\rm rest}(1-(v_s(r)/c))^{-1} >0$.

Another branch  $E_{\rm rest}=(v_s(r)+c)p_r$ in Fig.~\ref{TunnelingTraj}
contains two disconnected pieces describing the particle propagating from
the horizon in two opposite directions:
\begin{eqnarray}
r>r_h~:~p_r(r)= {E_{\rm rest}\over c+v_s(r)}  ~, ~ E_{\rm com}(r)=
cp_r(r)>0
\label{TunnelingTrajectory2}
\\
r<r_h~:~p_r(r)= {E_{\rm rest}\over c+v_s(r)}  ~,  ~E_{\rm
com}(r)= c p_r(r)<0
\label{TunnelingTrajectory1}
\end{eqnarray}
 The Eq.(\ref{TunnelingTrajectory2}) describes  the
outgoing particles -- the particles propagating from the horizon to the
exterior. The energy of these particles is positive in both frames, comoving
and rest.  The Eq.(\ref{TunnelingTrajectory1}) describes the propagation
of particles from the horizon to the orifice  (or to the singularity). Though
for the rest frame observer the energy of these particles is positive, these
particles, which live  within the horizon, belong to the Minkowski vacuum
in the
comoving frame.

The classical trajectory in
Eqs.(\ref{TunnelingTrajectory2},\ref{TunnelingTrajectory1}) is thus
disrupted at
the horizon. There is however a quantum mechanical transition between the
two pieces of the branch: the quantum tunneling. The tunneling amplitude can be
found in semiclassical approximation by shifting  the contour of integration to
the complex plane:
\begin{eqnarray}
w\sim  \exp (-2 S)~,\label{TunnelingProbability}\\
S={\bf Im}\int dr~ p_r(r)=  {\pi E_{\rm rest}\over | v_s'(r)|_{r=r_h}}~.
\label{TunnelingExponent}
\end{eqnarray}
This means that the wave function of any particle in the Minkowskii vacuum
inside the horizon contains an exponentially small part
describing the propagation from the horizon to infinity. This corresponds to
the radiation from the Minkowski vacuum in the presence of the event
horizon. The
exponential dependence  of the probability  on the quasiparticle energy
$E_{\rm rest}$ suggests that this radiation looks as thermal. The corresponding
temperature, the Hawking temperature, is
\begin{equation}
T_{\rm Hawking}= {\hbar | v_s'(r)|_{r=r_h} \over 2\pi}~.
\label{HawkingT}
\end{equation}

The radiation leads to the
quantum friction: the linear momentum of the flow decreases with time. This
occurs continuously until the superfluid Minkowski vacuum between the horizons
is completely exhausted and the superfluid state is violated. This leads to
the phase slip event, after which the number $N_3$ of circulation
quanta of superfluid velocity trapped by the torus is reduced. This process
will
repeatedly continue until the two horizons merge.

\subsection{Discussion.}

The above construction in Fig. \ref{2DBH}(b) allows us (at least in
principle) to
obtain the event horizon in the quasi-stationary regime, when the main
source of nonstationarity is the dissipation coming from the Hawking
radiation. As for the practical realization, there are, of course, many
technical problems to be solved. On the other hand, if the black hole
analog can in principle exist in condensed matter as the quasi-stationary
object,
its prototype -- the real black hole -- can also exist, at least in principle
(though it is not so easy to find the scenario of how this object can be
obtained
from the gravitational collapse of matter
\cite{Evanescent}).

It appears that
in some range of the energies of Hawking-radiated particles, close to the
"Planck" scale, where the fermionic spectrum becomes "nonrelativistic", the
Hawking radiation does not exist any more. The particles which are radiated in
the fully relativistic case, will be now Andreev scattered back to the black
hole. Thus both partners (particle and hole) of the Hawking radiation will
remain
within the horizon, so that the particle creation in high gravity field will
disturb the quantum vacuum inside the horizon without any radiation outside. In
principle the pair creation inside the horizon can be more important for the
dissipation in the superfluid 3He film under discussion than the Hawking
radiation.

\section{Rotating vacuum.}

\subsection{Unruh effect.}

The  body moving in the vacuum  with linear  acceleration $a$ is believed to
radiate the thermal spectrum with the Unruh temperature $T_U=\hbar a/2\pi c$
\cite{Unruh1}. The comoving observer sees the vacuum as a thermal bath
with $T=T_U$, so that the matter of the body gets heated to $T_U$ (see
references
in \cite{Audretsch}). Linear motion at constant proper acceleration
(hyperbolic motion) leads to velocity arbitrarily close to the speed of
light. On the other hand uniform circular motion features constant
centripetal acceleration while being free of the above mentioned pathology
(see the latest references in
\cite{RotatingQuantumVacuum,Leinaas,OrbitingUnruh,Barber}). The latter
motion is stationary in the rotating frame, which is thus a convenient
frame for study of the radiation and thermalization effects for uniformly
rotating body.

\subsection{Zel'dovich-Starobinsky effect.}

Zel'dovich \cite{Zeldovich1} was the first who predicted that the rotating
body (say, dielectric cylinder) amplifies those electromagnetic modes which
satisfy the condition
\begin{equation}
 \omega - L \Omega<0 ~.
\label{ZeldovichCondition}
\end{equation}
Here $\omega$ is the frequency of the mode, $L$ is its azimuthal quantum
number, and $\Omega$ is the angular velocity of the rotating cylinder.
This amplification of the incoming radiation is referred to as
superradiance
\cite{BekensteinSchiffer}. The other aspect of this phenomenon is that due
to quantum effects, the cylinder rotating in quantum vacuum
spontaneously emits the electromagnetic modes satisfying
Eq.(\ref{ZeldovichCondition})
\cite{Zeldovich1}. The same occurs for any rotating body, including the
rotating black hole \cite{Starobinskii}, if the above condition is satisfied.

Distinct from the linearly accelerated body, the radiation by a rotating
body does not look thermal. Also, the rotating observer does not see the
Minkowski vacuum as a thermal bath.  This means that the matter of the body,
though excited by interaction with the quantum fluctuations of the Minkowski
vacuum, does not necessarily acquire an intrinsic temperature  depending
only on
the angular velocity of rotation. Moreover the vacuum of the rotating frame
is not
well defined because of the ergoregion, which exists at the distance
$r_e=c/\Omega$ from the axis of rotation.

The problems related to the response of the quantum system  in its ground
state to rotation\cite{RotatingQuantumVacuum}, such as radiation by the
object rotating in vacuum
\cite{Zeldovich1,Zeldovich2,Starobinskii,BekensteinSchiffer} and  the
vacuum instability caused by the existense of ergoregion
\cite{QuantumErgoregionInstablity}, etc., can be simulated in superfluids,
where
the superfluid ground state plays the part of the quantum vacuum. The
quantum friction due to spontaneous emission of phonons in superfluid
$^4$He and Bogoliubov fermions in superfluid $^3$He-B has been discussed
in \cite{CalogeracosVolovik}. Here we extend an analysis to the radiation
of quasiparticles in $^3$He-A.

\subsection{Cylindrical geometry.}

Let us consider a cylinder of radius $R$ rotating with angular
velocity $\Omega$ in the (infinite) superfluid liquid.
In this situation there are again two important reference frames. One of
them is the laboratory frame.  The
energy of quasiparticles is not well determined in the laboratory frame,
because, if the rotating body is not the perfect axysymmetric cylinder, it
produces the time dependent perturbations of the liquid. However at
distances far enough from the surface of the cylinder the influence of the
rotating cylinder on the liquid can be neglected. In this
approximation the superfluid vacuum is in rest with respect to the
laboratory frame. This means that the superfluid velocity
${\bf v}_s=0$ in the laboratory frame. The quasiparticle energy in this
frame is that as in the comoving frame of previous Section. It is
$E_{\rm com}=cp$ if we consider superfluid $^4$He, or
the Eq.(\ref{EnergyComovingFrame}),  if we consider a pure 2D motion of
$^3$He-A with
$\hat{\bf l}=\hat{\bf z}$. Such energy spectrum corresponds to the
effective Minkowski metric of flat space
 \begin{equation}
ds^2=- c^2 dt^2   + r^2d\phi^2+dr^2  + a^2dz^2
 ~.
\label{Intervallz}
\end{equation}
Here $a=1$ for  isotropic $^4$He and $a= c/v_F$ for anisotropic $^3$He-A.

\subsection{Conical texture with negative angle deficit.}

In the 3D case of $^3$He-A there can be another effective
metric far from the body.  It can be caused by the normal boundary
condition on the $\hat{\bf l}$-vector, which prefers the radial
orientation of $\hat{\bf l}$-vector.  In the case of the radial
distribution,
$\hat{\bf l}=
\hat{\bf r}$,  the effective metric for the
quasiparticles moving outside the cylinder follows from
Eq.(\ref{GravFieldInAPhase}):
 \begin{equation}
ds^2=- c^2dt^2  +dz^2 + r^2d\phi^2+{c^2\over v_F^2}dr^2
 ~.
\label{Interval2}
\end{equation}
Such an effective space is conical:  The space outside the
cylinder is flat, but the proper length  of the circumference of radius
$r$ around the cylinder is not equal to $2\pi r$. In the
relativistic theories such conical metric can arise outside the
local cosmic strings, where there is an angle deficit. In our case the
length  of the
circumference is $2\pi r v_F/c $, which is much larger than $2\pi r$: This
effective space
exhibits a "negative angle  deficit" (for details see
Ref.\cite{GravityOfMonopole}).

\subsection{Rotating frame.}

Since the system far
outside the rotating cylinder is not disturbed by the rotation of the
body,  the system will remain to be in or close to the Minkowski vacuum
(or vacuum in the conical space in the case of radial $\hat{\bf
l}$-vector) as viewed in the laboratory frame. However this does not hold
in the region ajacent to the cylinder, where the superfluid
velocity field is disturbed and time-dependent. Also the energy in the
laboratory frame is not well determined because of the time-dependent
perturbations.

The energy is well determined in the frame corotating with the
cylinder. In the corotating frame the cylinder is at rest, and thus the
perturbations caused by rotation are stationary. The metric in the
coorotating frame is simplest far outside the rotating body, where the
superfluid velocity in the corotating frame is ${\bf
v}_s=-\vec\Omega\times {\bf r}$. Substituting this
${\bf v}_s$ into Eq.(\ref{GravFieldInAPhase})  one
obtains the interval
$ds^2=g_{\mu\nu}dx^\mu dx^\nu$,  which determines the propagation of
quasiparticles in the corotating frame:
 \begin{equation}
ds^2=-(c^2-\Omega^2r^2)dt^2 - 2\Omega r^2d\phi dt +
r^2d\phi^2+dr^2 +a^2dz^2
 ~.
\label{IntervalRotation}
\end{equation}
Here again $a=1$ for phonons in isotropic $^4$He and $a= c/v_F$ for
fermionic quasiparticles in anisotropic
$^3$He-A with $\hat{\bf l}=\hat{\bf z}$.
These metrics correspond to the interval in the rotating frame discussed
in relativistic theories.

It is convinient to write the quasiparticle spectrum in this frame in two
different approximations. In classical description one has $E_{\rm
corotating} =E_{\rm com}({\bf p}) +{\bf p}\cdot {\bf v}_s$.  In the other
description, an azimuthal motion of quasiparticles is quantized in terms of
the angular momentum $L=rp_\phi$, while the radial is still treated in the
quasiclassical approximation. In this case the energy spectrum of phonons
in the corotating frame is
\begin{equation}
E_{\rm corotating} =cp +{\bf p}\cdot {\bf v}_s= c\sqrt{ {L^2\over r^2} +
p_z^2 +  p_r^2} - \Omega L
 ~,
\label{QuasiclassicalPhononSpectrum}
\end{equation}
and the energy spectrum of the Bogoliubov fermions
\begin{equation}
E_{\rm corotating}({\bf p})=\pm\sqrt{{c^2 \over r^2}L^2 + c^2p_r^2  +
v_F^2(p_z\mp p_F)^2} -\Omega L
 ~.
\label{RotBogolonSpectrum}
\end{equation}

\subsection{Ergoregion in superfluids.}

The radius $r_e=c/\Omega$, where $g_{00}=0$ in the
Eq.(\ref{IntervalRotation}), marks the position of the ergoplane. In the
ergoregion, i.e. at  $r>r_e=c/\Omega$, the energy of quasiparticle
in Eq.(\ref{RotBogolonSpectrum}) can be negative for any $ L\geq
1$. This means that in the ergoregion the Minkowski vacuum is not the
vacuum for the corotating observer.
Situation is similar to the case discussed in previous Section. However
the relevant superfluid velocity is azimuthal now instead of radial. As a
result there is an ergoplane instead of a horizon.

From the point of view of the corotating observer, the Minkowski vacuum
is unstable  towards the filling of the negative energy states in the
ergoregion, which means the radiation of the quasiparticles from the
rotating cylinder. However we need a real process, which leads to such
radiation. This radiation can be caused only by the interaction
between the superfluid Minkowski vacuum and the rotating object.

Let us consider the slow rotations $\Omega R \ll c$. In this case the
linear velocity of the cylinder at the surface of the cylinder, $\Omega
R$, is much smaller than the Landau critical velocity for nucleation of
quasiparticles,
$v_{Landau}=c$. Thus quasiparticles cannot be nucleated near the surface of
cylinder.  The
ergoregion, where $|v_s|=\Omega r >c$ and quasiparticles can be nucleated,
is far from the cylinder, $r_e\gg R$. The
interaction with the cylinder, which produces the matrix element for the
radiation, is very small. In this situation the most effective mechanism of
the quasiparticle radiation is the tunneling of quasiparticles from the
liquid ajacent to the surface of the rotating body, which plays the part
of the rotating detector, to the ergoregion.

\begin{figure}[!!!t]
\begin{center}
\leavevmode
\epsfig{file=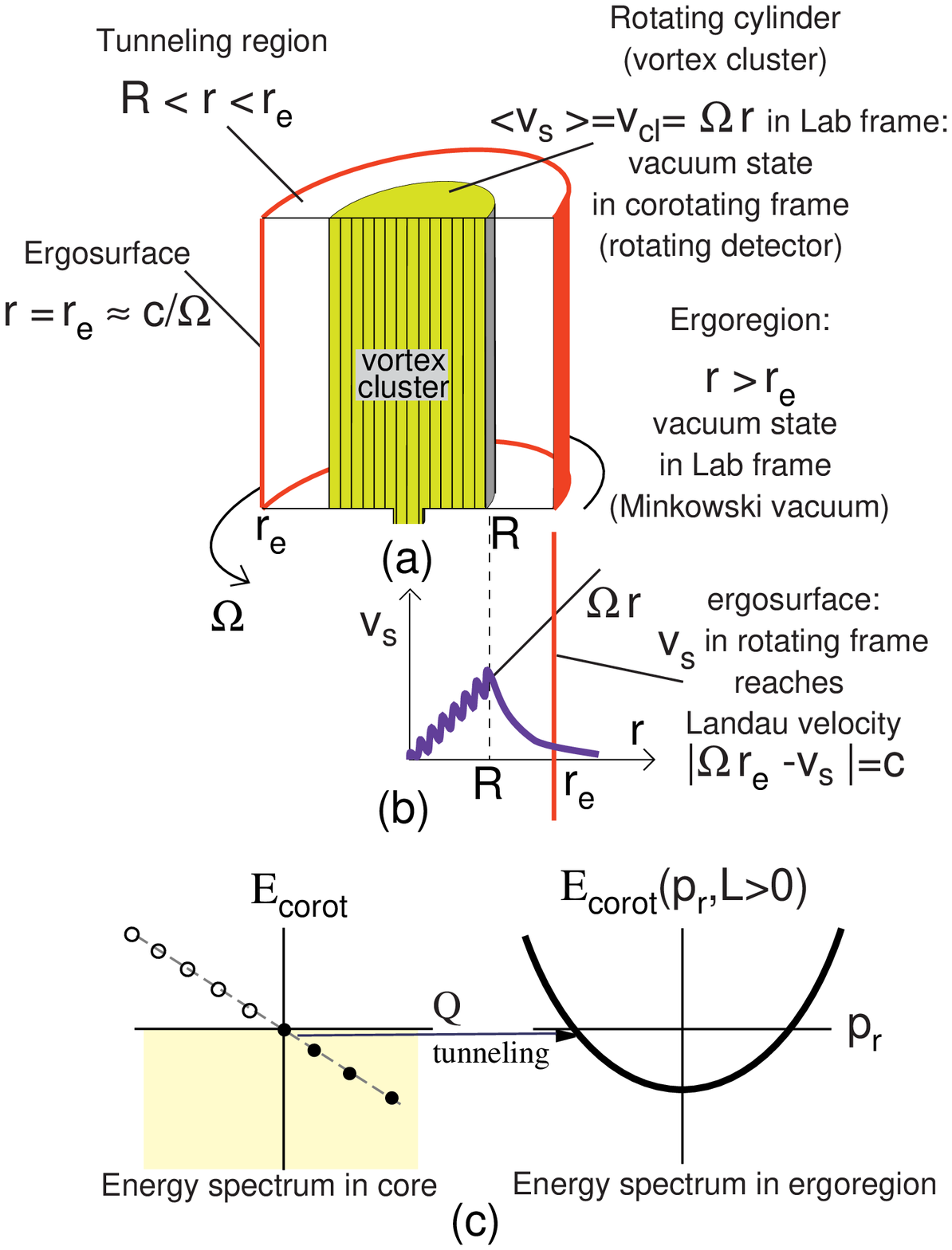,width=0.9\linewidth}
\caption[Ergoregion]
    {(a) Rotating body -- rigidly rotating cluster of vortices.
Within the cluster, at
$r<R$,  the average superfluid velocity equals the
velocity $v_{\rm cl}$ of the solid body rotation of the cluster.
Superfluid state within the cluster plays the part of rotating detector:
this is the vacuum state in the corotating frame. Far outside of the
cluster, where velocity of the superfluid is zero in the Lab frame, the
system is in Minkowski vacuum. (b) Distribution of the superfluid velocity
in the laboratory frame. At $r=r_e\approx c/\Omega$ the superfluid
velocity in the corotating frame reaches the "speed of light" -- the
Landau critical velocity $v_{Landau}=c$. At $r>r_e$ there is an
ergoregion in the corotating frame, where the quasiparticle
negative energy states are empty. (c) Tunneling of quasiparticles from
the vacuum state of the rotating detector to the "Minkowski" vacuum in
the ergoregion.  This produces radiation from the rotating body
(from vortex cluster) and excitation of the rotating detector (the fermion
zero modes in the cores of vortices) }
\label{Ergoregion}
\end{center}
\end{figure}

\subsection{Rotating detector.}

The simplest rotating detector is the fermionic system, which is rigidly
connected to the rotating body. In superfluids the rotating body can be
effectively substituted by the rigidly rotating cluster of quantized
vortices (see Fig.~\ref{Ergoregion}(a)). Such clusters, that experience
the solid-body rotation, are experimentally investigated in both phases of
superfluid
$^3$He (see e.g. \cite{Cluster}).
The vortex cluster rotating in the infinite superfluid liquid represents
the  vacuum in the corotating frame. However this vacuum state is not
extended to the exterior of the cluster, where one has the analogue of
Minkowski vacuum. That is why such state is not equilibrium, but it is
quasistationary metastable state  which can live for a macroscopically long
time. At $T=0$ the
dominating mechanism of the relaxation of angular velocity $\Omega$ of
the cluster is the process of the radiation of quasiparticles.

Fig.~\ref{Ergoregion}(b) shows the distribution of superfluid velocity
$v_s$ in the laboratory frame. Within the vortex cluster, i.e. at $r<R$,
the superfluid velocity, being averaged over the vortices, follows the
velocity of the solid body rotation of the cluster: i.e.
$<{\bf v}_s>=\vec\Omega\times {\bf r}$ in the laboratory frame and thus
$<{\bf v}_s>=0$ in the  frame corotating with the cluster. Outside the
cluster the superfluid velocity decays as $N\kappa/2\pi r$, where $N$ is
number of vortices in the cluster and $\kappa$ is superfluid circulation
around individual vortex.

Within the cluster one has $<{\bf v}_s>=0$ in the corotating frame,
so one can expect that the  spectrum of quasiparticles, which live within
the cluster, is the same as the energy spectrum  in the frame comoving with
the superfluid velocity, $E_{\rm corotating}=E_{\rm com}$. In other words
this spectrum has no $-\Omega L$ shift of the energy levels, as distinct
from quasiparticle spectrum far from the body,
Eqs.(\ref{QuasiclassicalPhononSpectrum},\ref{RotBogolonSpectrum}), which
is measured in the same corotating frame.
This is not exactly true, since the equation $<{\bf v}_s>=0$ does not
imply that
${\bf v}_s=0$ locally: it essentially depends on the position in the vortex
lattice.

The main feature is nevertheless preserved, when one considers the
spectrum of quasiparticles that live in the vortex core. Their energy
spectrum in the corotating state does not depend on rotation velocity
$\Omega$:  $E_{\rm corotating}=-\omega_0(p_z) Q$, where $Q$ is the
generalized angular momentum; $\omega_0(p_z)$ is the so called minigap,
which depends on the core structure. This branch crosses zero energy level
as a function of $Q$, if one considers this quantum number as continuous,
and thus represents the fermion zero mode (see
left part of Fig.~\ref{Ergoregion}(c)). The energy spectrum of such chiral
fermions, that live in the vortex core, has been first calculated by
Caroli,  de Gennes  and  Matricon for the Abrikosov vortex in
$s$-wave superconductors \cite{Caroli}; on the relation between the number
of fermion zero modes and the winding number of the vortex see
\cite{IndexVortices}. For us it is important that for vortices in
$^3$He the quantum number $Q$ is integer and thus one has the states in
the detector with zero energy in the corotating frame.

\subsection{Radiation to the ergoregion.}

The radiation of
Bogoliubov quasiparticles can be considered as the process in which
the particle from the zero energy state in the detector, $E_{\rm
corotating}=0$, tunnels to the scattering state at the ergoplane,  where
also its energy is
$E_{\rm corotating}=0$ (Fig.~\ref{Ergoregion}(c)). In the quasiclassical
approximation the tunneling probability is
$e^{-2S}$, where at $p_z=\pm p_F$ and  $\Omega R \ll c$:
\begin{equation}
S={\rm {Im}} \int dr~p_r(r;E=0)=L \int^{r_e}_{R}dr~\sqrt{{1\over r^2}-
{1\over r_e^2}}\approx  L  \ln {r_e\over R}
 ~.
\label{Tunneling}
\end{equation}
Thus all the particles with
$L>0$ are radiated, but the radiation probability is smaller for higher
$L$:
\begin{equation}
w \propto e^{-2S}=\left({R\over r_e}\right)^{2L}=\left({\Omega R\over
c}\right)^{2L} = \left({\omega R\over
cL}\right)^{2L}~,~\Omega R \ll c
 ~.
\label{FermionTunnelingProbability}
\end{equation}
Here $\omega=\Omega L$  is energy (frequency) of the radiated
quasiparticles in the laboratory frame.

If $c$ is substituted by the speed of light,
Eq.(\ref{FermionTunnelingProbability}) is proportional to the superradiant
amplification of the electromagnetic  waves by rotating  dielectric
cylinder derived by Zel'dovich
\cite{BekensteinSchiffer,Zeldovich2}.

Since each radiated fermion carries the angular momentum $L$, the
vortex cluster  rotating in superfluid vacuum at $T=0$  is loosing its
angular momentum and thus experiences  the quantum rotational
friction. The radation also leads to excitation of the detector matter.
In principle the radiation can occur without excitation of the
 detector vacuum, via direct interaction of the particles in the Minkowski
vacuum with the rotating body (cluster), but the contribution of this
process the  quantum friction is smaller \cite{CalogeracosVolovik}.

\subsection{Discussion.}

The rotational friction experienced by the body rotating in superfluid
vacuum at
$T=0$, is caused by the  spontaneous quantum emission of the quasiparticles
from
the rotating object to the "Minkowski" vacuum in the ergoregion. The
emission is not
thermal and depends on the details of the interaction of the radiation with the
rotating body.  In the quasiclassical approximation it is mainly determined
by the tunneling exponent, which can be approximately characterized by the
effective temperature
$T_{\rm eff}\sim \hbar \Omega (2/\ln (c/\Omega R))$. The vacuum friction of the
rotating body can be observed only if the effective temperature exceeds the
temperature of the bulk superfluid, $T_{\rm eff}>T$. For the body rotating with
$\Omega=10^3$rad/s,
$T$ must be below $10^{-8}$K. However, high rotation velocity  can be
obtained for clusters. The cluster containg two vortices
rotate around their center of mass with
$\Omega=\kappa/4\pi R^2$, where
$R$ is the radius of the  circular orbit. If the radius $R$ is of order
of superfluid coherence length, the effective temperature can reach
$10^{-4}$K.

The process discussed in this Section occurs only if there is an ergoplane
in the
rotating frame. If the superfluid is contained in a finite  external
cylinder of radius
$R_{\rm ext}>R$, this process occurs only at high enough rotation velocity,
$r_e(\Omega)=c/\Omega < R_{\rm ext}$, when the ergoplane is
within the superfluid.   On the instability of the ergoregion in
quantum vacuum towards emission see also in
Ref.\cite{QuantumErgoregionInstablity}.

If  $r_e(\Omega)> R_{\rm ext}$ and ergoregion is not present, then the
interaction between the coaxial cylinders via the vacuum
fluctuations becomes the main mechanism for dissipation. This causes the
dynamic Casimir forces between the walls moving laterally  (see Review
\cite{Kardar}). As in \cite{Kardar} the nonideality of the cylinders, i.e.
violation
of the rotational symmetry by the body, is the necessary condition for quantum
friction.

\section{Discussion}

In the above examples of the nontrivial space, the effective gravitational
field acts
as a fixed external field. The dynamics of this field has not been
discussed here.
In most cases the effective gravity field do not obey
the Einstein equations. This is the main drawback of superfluid $^3$He-A:
Since the
Fermi points in $^3$He-A are too far apart from each other, the dynamical
equations
for the gravitational field are clumsy. Nevertheless the above textures
allow us to
simulate many phenomena related to quantum vacuum in the presence of the strong
gravity field. This is because many properties of the quantum vacuum in
curved space,
which are determined by the geometry, do not depend on the dynamical origin
of the
geometry.  For example, it is well known that the Hawking radiation is a purely
kinematic effect and occurs in any geometry, if it exhbits an event horizon
\cite{Visser1999}. That is why the $^3$He-A quantum vacuum is a right
object for
simulation of many aspects of physics of vacuum in a curved space. For
example what is
the effect of the breakdown of the Lorentz invariance at higher energy on
Hawking
radiation. The entropy of the black hole can be also investigated using the
above
model, since the microstates within the horizon are well determined and (at
least in
principle) are completely known in the whole energy range including the
"transPlanckian" region. All these are different aspects of the problem
of stability of the vacuum in strong gravitational and other fields. The
superfluids
provide many examples of the instability of the superfluid vacuum, and thus
allow us
to investigate different mechanisms of relaxation of the physical vacuum --
the ether.

\vfill\eject

\end{document}